\documentclass[sigconf]{acmart}
\AtBeginDocument{%
  }

\usepackage{multirow}
\usepackage{hyperref}
\usepackage{enumitem}
\usepackage{graphicx}
\usepackage{subcaption}
\usepackage{graphicx}
\usepackage{array}
\usepackage{longtable}
\usepackage{lscape}
\usepackage{makecell}
\usepackage{orcidlink}
\usepackage{balance}

\usepackage{xcolor}

\graphicspath{ {./fig/}  } 

\copyrightyear{2025}
\acmYear{2025}
\setcopyright{rightsretained}
\acmConference[CSCW Companion '25]{Companion of the Computer-Supported Cooperative Work and Social Computing}{October 18--22, 2025}{Bergen, Norway}
\acmBooktitle{Companion of the Computer-Supported Cooperative Work and Social Computing (CSCW Companion '25), October 18--22, 2025, Bergen, Norway}\acmDOI{10.1145/3715070.3749232}
\acmISBN{979-8-4007-1480-1/2025/10}

\begin{document}
% \maketitle

\title{Linking Facial Recognition of Emotions and Socially Shared Regulation in Medical Simulation}

\author{Xiaoshan Huang}
\affiliation{%
  \institution{McGill University}
  \city{Montréal}
  \country{Canada}}
\email{xiaoshan.huang@mail.mcgill.ca}
\orcid{0000-0002-2853-7219}

\author{Tianlong Zhong}
\affiliation{%
  \institution{Nanyang Technological University}
  \country{Singapore}}
\email{tianlong001@e.ntu.edu.sg}
\orcid{0000-0002-9467-2881}

\author{Haolun Wu}
\affiliation{%
  \institution{Stanford University}
  \city{Palo Alto}
  \country{USA}
}
\email{haolunwu@cs.stanford.edu}
\orcid{0000-0002-8206-4969}

\author{Yeyu Wang}
\affiliation{%
 \institution{University of Wisconsin-Madison}
 \city{Madison}
 \state{Wisconsin}
 \country{USA}}
 \email{ywang2466@wisc.edu}

\author{Ethan Churchill}
\affiliation{%
 \institution{McGill University}
 \city{Montréal}
 \state{Quebec}
 \country{Canada}}
 \email{ethan.churchill@mail.mcgill.ca}

\author{Xue Liu}
\affiliation{%
  \institution{McGill University \& Mila-Quebec Artificial Intelligence Institute}
  \city{Montréal}
  \country{Canada}}
\email{xue.liu@cs.mcgill.ca}

\author{David Williamson Shaffer}
\affiliation{%
 \institution{University of Wisconsin-Madison}
 \city{Madison}
 \state{Wisconsin}
 \country{USA}}
 \email{dws@education.wisc.edu}
 
\renewcommand{\shortauthors}{Xiaoshan Huang et al.}
%% If needed use a foot or authornote to identify equal contribution

\begin{abstract}
Computer-supported simulation enables a practical alternative for medical training purposes. This study investigates the co-occurrence of facial-recognition-derived emotions and socially shared regulation of learning (SSRL) interactions in a medical simulation training context. Using transmodal analysis (TMA), we compare novice and expert learners’ affective and cognitive engagement patterns during collaborative virtual diagnosis tasks. Results reveal that expert learners exhibit strong associations between socio-cognitive interactions and high-arousal emotions (surprise, anger), suggesting focused, effortful engagement. In contrast, novice learners demonstrate stronger links between socio-cognitive processes and happiness or sadness, with less coherent SSRL patterns, potentially indicating distraction or cognitive overload. Transmodal analysis of multimodal data (facial expressions and discourse) highlights distinct regulatory strategies between groups, offering methodological and practical insights for computer-supported cooperative work (CSCW) in medical education. Our findings underscore the role of emotion-regulation dynamics in collaborative expertise development and suggest the need for tailored scaffolding to support novice learners’ socio-cognitive and affective engagement.

\end{abstract}


%% CCS concept
\begin{CCSXML}
<ccs2012>
<concept>
<concept_id>10010405.10010489.10010492</concept_id>
<concept_desc>Applied computing~Collaborative learning</concept_desc>
<concept_significance>500</concept_significance>
</concept>
<concept>
<concept_id>10010405.10010489.10010491</concept_id>
<concept_desc>Applied computing~Interactive learning environments</concept_desc>
<concept_significance>500</concept_significance>
</concept>
<concept>
<concept_id>10003120.10003121.10011748</concept_id>
<concept_desc>Human-centered computing~Empirical studies in HCI</concept_desc>
<concept_significance>300</concept_significance>
</concept>
</ccs2012>
\end{CCSXML}

\ccsdesc[500]{Applied computing~Collaborative learning}
\ccsdesc[500]{Applied computing~Interactive learning environments}
\ccsdesc[300]{Human-centered computing~Empirical studies in HCI}


\keywords{Facial Recognition, Socially-Shared Regulation, Transmodal Analysis, Medical Simulation, Emotions, Multimodal Learning Analytics}

%\renewcommand\footnotetextcopyrightpermission[1]{}
%\pagestyle{plain}  % removes headers/footers like ACM reference format

\maketitle 

\vspace{-1mm}
\section{Introduction} 
\begin{figure*}
    \centering
    \includegraphics[width=1.0\linewidth]{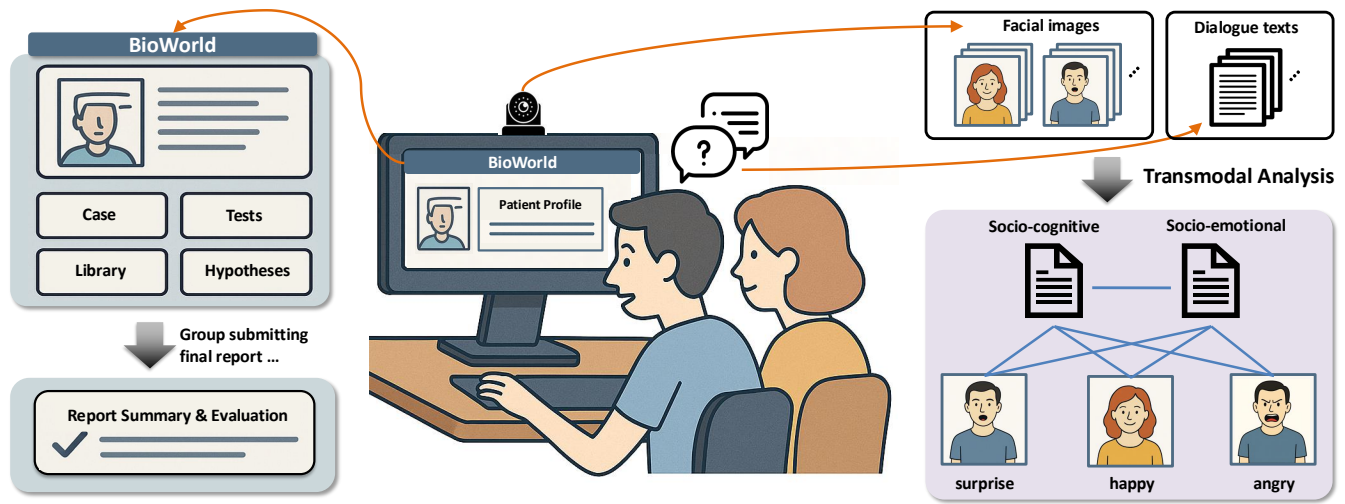}
    \caption{BioWorld Interface and Data Retrieval Processes in the Virtual Medical Diagnostic Task}
    \label{fig:bioworld_transmodel}
    \Description{The left part illustrates the medical simulation interface for dyads to solve a patient case. The middle part shows the cooperative medical simulation setting. And the right part shows how multi-modal data, in our context, learners' facial expressions and their dialogues, are extracted and analyzed.} 
    \vspace{1mm}
\end{figure*}
Medical simulation provides a practical alternative for training real-time diagnostic skills, which are often time-consuming and costly to practice in real clinical settings. Computer-supported collaborative problem-solving further enhances learning by allowing practitioners to exchange and synthesize ideas, and refine decision-making collectively~\cite{zhu2023synthesis}. However, while medical professionals are trained to manage their reactions in patient-facing scenarios, little is known about how learners interact emotionally and cognitively in virtual diagnostic environments.

Understanding these interaction patterns and their associated emotional experiences through verbal and non-verbal cues are critical for designing effective educational technologies~\cite{breideband2023community}. Prior research highlights the pivotal role of socially shared regulation of learning (SSRL,~\cite{panadero2015socially}) on performance in collaborative learning settings, which involves meta-cognitive, cognitive, emotional, and motivational processes. Yet, the dynamic relationship between SSRL and real-time emotional responses—especially in medical simulation—remains underexplored. Facial recognition technology offers a promising tool to capture subtle emotional cues during diagnostic tasks, providing insights into how novice and expert learners differ in their collaborative approaches.

This study investigates how learners' SSRL strategies and emotional responses revealed through facial expressions intertwine during virtual medical diagnosis, with a focus on comparing novice and expert behaviors. By analyzing these interactions, we aim to inform the development of adaptive simulation platforms that optimize scaffolding and learner engagement.

\section{Related Work}
We review prior work on socially shared regulation in computer-based learning environments and the use of facial recognition in medical simulation.

\subsection{Socially Shared Regulation in Computer-Based Learning Environments}
SSRL strategies have been found to significantly influence learners' collaborative experiences in computer-based learning environments. Higher engagement in SSRL strategies is observed in high-performance teams where team members can more flexibly adapt SSRL strategies in front of challenges~\cite{malmberg2015promoting}.

Recent research started viewing SSRL interactions through micro lens. Meta-cognitive interactions facilitate the monitoring of cognitive processes and support actionable insights. Socio-cognitive interactions encompass team members' planning, goal-setting, evaluation of strategy effectiveness, and reflective practices. Socio-emotional interactions involve the explicit expression of emotional states during tasks, while socio-motivational interactions foster team cohesion, confidence in task completion, and satisfaction with collaborative efforts.

However, few studies have explored how these SSRL strategies are associated with user experience in medical simulation.

% \vspace{-1mm}
\subsection{Facial Recognition in Medical Simulation}
Facial recognition technology (FRT) is being adopted more widely in various fields ~\cite{ullstein2024mapping}, including education-based contexts such as medical simulation. For instance, a previous study found FRT effectively evaluate nursing students' emotional responses during clinical simulations, revealing areas of proficiency and difficulty—often marked by expressions of anger or other negative emotions~\cite{mano2019using}. This method enhances understanding of learners' engagement and receptiveness in simulated learning environments.

An integrative review highlighted that medical simulations often induce stress and fear in residents~\cite{madsgaard2022rollercoaster}, primarily due to feelings of unpreparedness, uncertainty about treatment, and evaluation anxiety~\cite{groot2020simulation,ko2020debriefing}. Previous studies have found that stress and anxiety during simulations do not always result in adverse effects on learning outcomes~\cite{ko2020debriefing}; and increased anxiety and stress can be a driver for learning and were associated with increased preparedness for the simulations~\cite{groot2020simulation}. However, emotional responses varied significantly, underscoring a gap in detailed knowledge about learners' affective states during simulations and inconsistent findings in existing literature~\cite{fraser2019temporal,schlairet2015cognitive}. Exploring medical residents' affective states through FRT could provide us better understanding on their learning processes especially in teamwork setting where team members take important roles in decision-making.

In this study, our core interest is to map out the co-occurrence of residents' emotions from facial recognition and their SSRL interactions in a medical simulation teamwork task, specifically we aim at answering the following research questions (RQs): RQ1a. What are the co-occurrences of emotion and SSRL behaviors during students' virtual diagnosis? RQ1b. Are there any differences in the co-occurrences of emotion captured by facial expressions and SSRL behaviors between novice and expert learners? 

\begin{figure*}[t]
    \centering
    \includegraphics[width=1.0\linewidth]{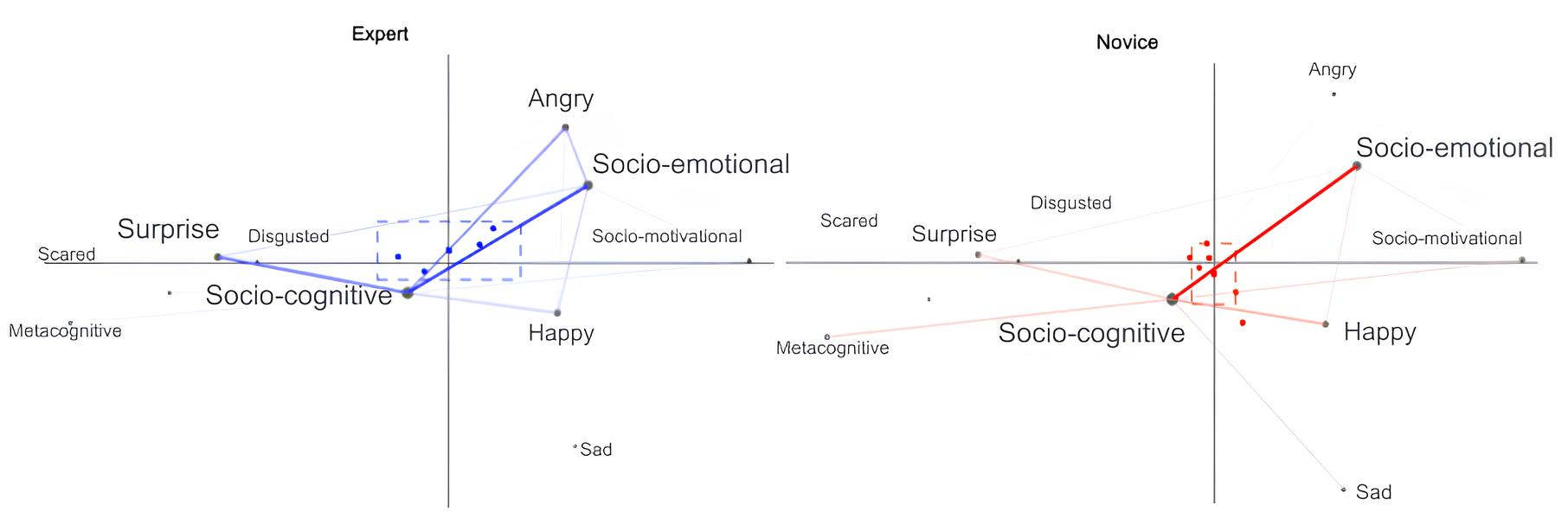}
    \caption{Co-Occurrences of Facial Recognition of Emotions and SSRL Interactions in the Expert and Novice Groups}
    \label{fig:tma_allgroups}
    % \vspace{-2mm}
    \Description{The left graph (in blue) represents the co-occurrences of expert learners with more years of residency experience and the right graph (in red) represents the co-occurrences of novice learners with less years of residency experience. The size of the dots and width of the lines represent the strength of the connection between two events.}
\end{figure*}

\begin{figure}[t]
    \centering
    \includegraphics[width=1.0\linewidth]{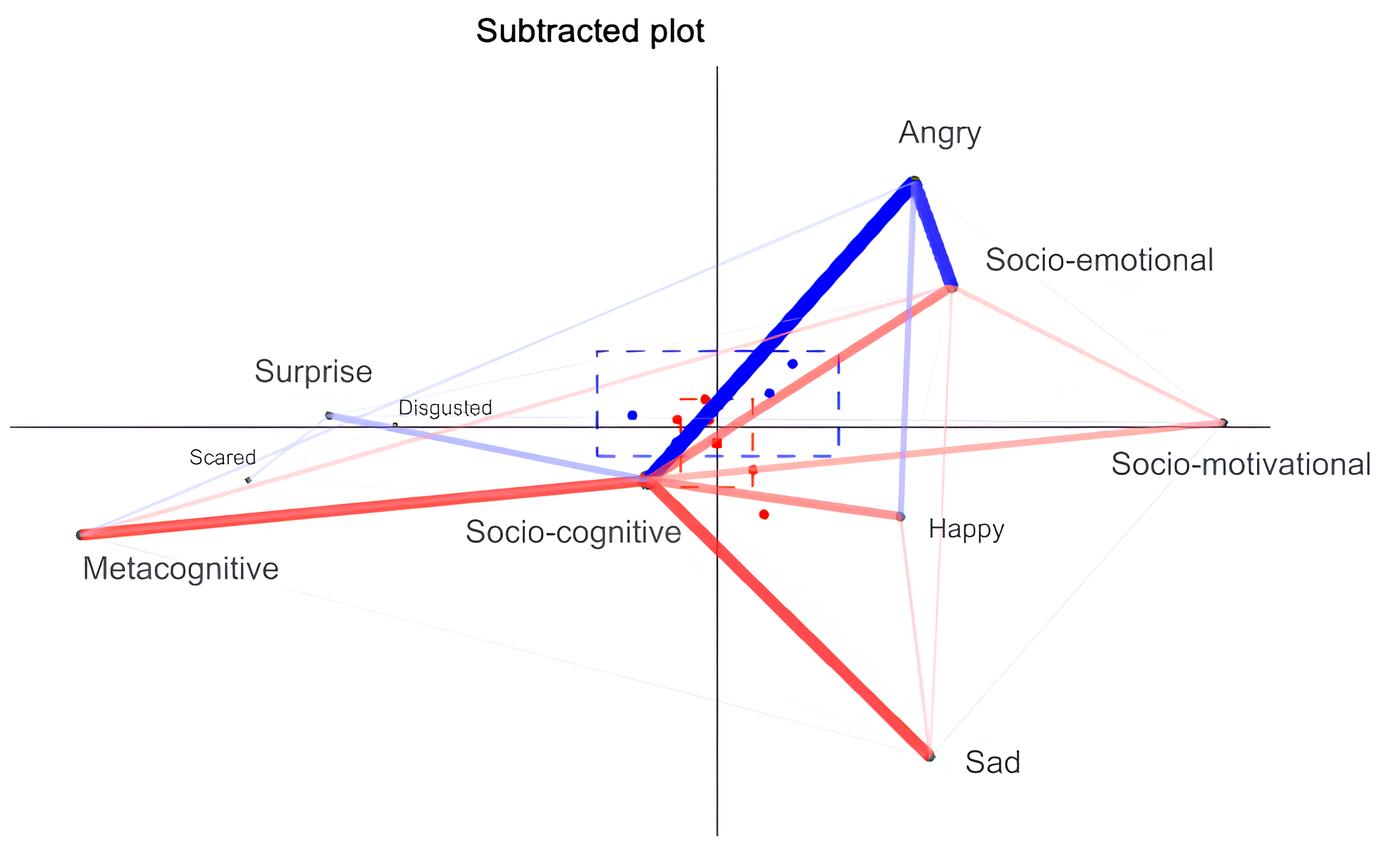}
    \caption{The Subtracted Network of Two Levels of Expertise}
    \label{fig:Substracted}
    \Description{Blue lines shows the connections that are stronger in expert learners (i.e., angry and socio-emotional, angry and socio-cognitive, surprise and socio-cognitive interactions); red lines shows the connections that are stronger in novice learners (e.g., meta-cognitive and socio-cognitive interactions, happy and socio-cognitive interactions)}
    \vspace{-1mm}
\end{figure}

% \vspace{-1mm}
\section{Method}
In this section, we introduce the computer-based learning environment and simulation task used in our empirical study, along with the data sources and analytic approach.
\subsection{Computer-Based Medical Simulation}
We use BioWorld \cite{lajoie2015modeling,huang2023exploring}, an Intelligent Tutoring System developed for medical students practicing medical reasoning as the computer-based problem solving platform in this study (see Figure \ref{fig:bioworld_transmodel}). The BioWorld interface has four functions: Case - to read and highlight key symptoms in the given patient profile; Tests - to order relevant tests for hypotheses examination; Library - to search specific terms or result for the patient's diagnosis; and Hypotheses - to rank current hypotheses with evidence collected and submit a final report. Learners can navigate through all functions at any time, and they will be given an expert solution after submitting their final report. 
\subsection{Participants and Multi-Modal Data Sources}
The university institutional review board (IRB) approved the research before the experiment took place (\#23-04-086) at the lab setting. Twelve medical residents paired into six dyads participated in this study. Among them, four male learners have more residency experience (at their 5th year of residency, Mean age = 29.50) and were first randomly paired into two dyads; the remaining eight female learners (Mean age = 28.13) were then randomized into four dyads. We then grouped participants into novice and expert learners groups according to their experience of residency. 

Participants were assigned to dyads and seated in front of the same screen to discuss about the diagnosis using BioWorld. The camera in front captured their facial expressions while solving the patient case. We use FaceReader 6.0 to extract and recognize their facial expressions into six basic emotions (i.e., happy, angry, surprise, sad, scared, and neutral). One participant’s facial recognition data was excluded from the study due to lack of consent. The microphones recorded learners' dialogue, which were further transcribed and coded into four types of SSRL interactions using an adapted coding scheme~\cite{huang2025makes}.

\subsection{Transmodal Epistemic Network Analysis
(T/ENA)}
Epistemic Network Analysis (ENA) is a method to represent and visualize connections among meaningful codes \cite{shaffer2016tutorial}. In educational research, ENA is especially valuable for identifying not just what students say or do, but also how their thinking unfolds through interconnected patterns with the contextual information in the learning processes. In this study, ENA operationalizes the patterns of collaborative learning as the co-occurrences of codes within the recent temporal contexts. That is, based on the transcripts, ENA constructs a network model based on the connections made within verbal and non-verbal codes for each individual in a group. Furthermore, ENA can subtract the mean networks between two groups, such as experts and novices, to manifest connection-making patterns in a collaborative discourse. In addition to visual comparisons, ENA can derive summative scores for individuals in a group and test differences in connection making through statistical tests.

However, ENA is usually used for unimodal learning analytics due to the constraint of a fixed-length sliding window. To account for unique temporal influences of verbal and non-verbal codes, we augmented ENA with Transmodal Analysis (TMA) to allow varied window sizes for cognitive and affective learning events \cite{shaffer2025transmodal}. Specifically, we specify the window of collaborative dialogue as 21.15 seconds, as an approximation of the recent temporal context, based on qualitative analysis \cite{ruis2019finding}; we specify the window of facial expressions to be an adaptive window corresponding to the synchronized dialogues. To conduct this T/ENA, we applied rENA and TMA packages in R.

\vspace{-1mm}
\section{Result}
This section presents the results of the transmodal network analysis, integrating participants’ facial emotion recognition data with their transcribed dialogues. We first illustrate the visualized outputs of these networks to highlight patterns of co-occurrence between emotional expressions and socially shared regulation of learning (SSRL) interactions. Then, we compare the differences between expert and novice learner groups by conducting a subtracted network analysis to reveal distinct emotional and regulatory dynamics across levels of expertise.

\subsection{RQ1a. Co-Occurrence of Facial Recognition and SSRL Interactions}

Figure \ref{fig:tma_allgroups} depicts the average ENA network of novice (in red) and expert (in blue) learners respectively. Both groups exhibit a strong co-occurrence between socio-cognitive and socio-emotional dimensions, suggesting a close relationship between residents’ cognitive and emotional states during virtual diagnosis. However, notable differences emerge in their emotional associations.

Expert learners demonstrate stronger links between socio-cognitive interactions and surprise and anger, both of which are high-arousal emotions. In contrast, novice learners show a higher correlation between happiness and socio-cognitive interactions, while their expressions of scare and anger remain relatively isolated.

% \vspace{-2mm}
\subsection{RQ1b. Co-Occurrence Differences between Novice and Expert Learners}
To further compare the difference of emotion and SSRL interactions co-occurrences, we constructed a subtracted graph (Figure \ref{fig:Substracted}). The results reveal that, compared to experts, novice learners exhibit stronger connections between socio-cognitive interaction and other SSRL interactions. Additionally, sadness and happiness are more frequently associated with socio-cognitive interactions among novices.

Conversely, experts display stronger links between socio-cognitive interactions and anger and surprise, while anger is also associated with socio-emotional interactions. This suggests distinct regulatory and emotional patterns between the two groups.

% \vspace{-2mm}

\section{Discussion}
Our study is the first to investigate the co-occurrence of facial-recognition-derived emotions and socially shared regulation of learning (SSRL) interactions in a medical simulation context. Using transmodal analysis, we mapped multi-modal data to uncover significant associations between residents’ expressed emotions and SSRL interaction types. Our findings not only validate this methodological approach but also provide insights into learners’ affective and cognitive engagement.

Experts’ facial expressions of surprise and anger were closely linked to socio-cognitive interactions. From an epistemic perspective~\cite{harley2015multi}, anger in this context may reflect concentration or frustration, as furrowed brows could indicate deep cognitive engagement rather than purely negative affect. Both surprise and anger are high-arousal emotions, suggesting that experts experience heightened engagement during socio-cognitive interactions.

In contrast, novices exhibited more happiness and sadness alongside socio-cognitive interactions. While happiness may indicate comfort, sadness could signal disengagement, hopelessness, or boredom—emotions indicating lower affective involvement. Furthermore, novices’ less distinct SSRL interaction patterns suggest they may struggle with task focus, possibly due to cognitive overload or difficulties navigating the virtual environment.

\section{Limitation and Future Works}
This study had a limited North American sample size and focused on a single diagnostic case. Future research should expand to diverse populations and examine how task complexity affects emotional states and SSRL dynamics. Despite these limitations, the study provides valuable methodological and practical insights for computer-supported cooperative work (CSCW) in medical training.

Methodologically, our findings demonstrate that multimodal learning analytics can serve as an effective framework for investigating cooperative dynamics in computer-supported medical education systems.

Practically, experts’ focused socio-cognitive engagement highlights their self-regulated learning efficiency, whereas novices’ fragmented patterns suggest a need for scaffolded support to enhance emotional and cognitive regulation. Future research could explore real-time affective feedback systems to optimize collaborative learning experiences. Additionally, human-computer interaction (HCI) researchers, learning scientists, and other stakeholders in medical training could collaborate on exploring the longitudinal development of regulation skills as novices transition to expertise.

\balance
\bibliographystyle{ACM-Reference-Format}
\bibliography{reference}

\end{document}